\documentclass[12pt]{article} 
\usepackage{epsfig}
\title{ Lepton-flavor violating decays as probes of quantum gravity?}
\author{ Z.~K.~Silagadze 
\vspace*{3mm} \\
Budker Institute of Nuclear Physics,  630 090,
Novosibirsk, Russia }
\date{}

\begin{document}
\large
\maketitle

\begin{abstract}
Lepton flavor violating decays  $Z \to \mu \tau$ and $J/\Psi, \Upsilon 
\to \mu \tau$ are considered. It is shown that these decays can reach
sizeable magnitudes if some specific lepton-flavor violating 4-fermion
operators are generated by low scale quantum gravity effects, or by some
other new physics at a TeV scale. 
\end{abstract}

\newpage

\newif\iftranslate
\translatefalse
%\translatetrue
\iftranslate
\section{Prologue -- a fragment from T.~D.~Lee's recollections \cite{1}}
``We found that if $\mu$-decay and $\mu$-capture were described by 
a four-fermion interaction similar to $\beta$-decay, all their coupling
constants appeared to be of the same magnitude. This began the universal 
Fermi interaction. We then went on to speculate that, in analogy with
electromagnetic forces, the basic weak interaction could be carried 
by a universal coupling through an intermediate heavy boson, which I
later called $W^\pm$ for weak. Naturally we went to our teacher,
Enrico Fermi, and told him of our discoveries. Fermi was extremely
encouraging. With his usual deep insight, he immediately recognized
the further implications beyond our results. He put forward the problem
that if this is to be the universal interaction, then there must be
reasons why some pairs of fermions should have such interactions, and some
pairs should not. For example, why does
$$p \rightarrow \hspace*{-6mm} / \hspace*{6mm}  e^+ \; + \; \gamma$$
\noindent and
$$p \rightarrow \hspace*{-6mm} / \hspace*{6mm} e^+ \; + \; 2\nu \; ?$$
\noindent A few days later, he told us that he had found the answer;
he then proceeded to assign various sets of numbers, +1, -1, and 0, to
each of these particles. This was the first time to my knowledge that
both the laws of baryon-number conservation and of lepton-number
conservation were formulated together to give selection rules. However,
at that time (1948), my own reaction to such a scheme was to be quite 
unimpressed: surely, I thought, it is not necessary to explain why
$p \rightarrow \hspace*{-6mm} / \hspace*{6mm}  e^+ \; + \; \gamma$, 
since everyone knows that the identity 
of a particle is never changed through the emission and absorption
of a photon; as for the weak interaction, why should one bother to
introduce a long list of mysterious numbers, when all one needs is
to say that only three combinations $(\bar n p)$, $(\bar e \nu)$, and
$(\bar \mu \nu)$ can have interactions with the intermediate boson. 
[Little did I expect that soon there would be many other pairs joining
these three.]

Most discoveries in physics are made because the time is ripe. If one 
person does not make it, then surely another person will do it at 
about the same time. In looking back, what we did in establishing the
universal Fermi interaction was a discovery of exactly this nature.
This is clear, since the same universal Fermi coupling observations
were made independently by at least three other groups, Klein, Puppi,
and Tiomno and Wheeler, all at about the same time. Yet Fermis's
thinking was of a more profound nature. Unfortunately for physics, his
proposal was never published. The full significance of these conservation
laws was not realized until years later. While this might be the first 
time that I failed to recognize a great idea in physics when it was 
presented to me, unfortunately it did not turn out to be the last.''
\fi

\section{Introduction} 
First published proposals about lepton number conservation are dated
back to 1953 \cite{1,2}. But only after a decade of intensive experimental 
research \cite{3}, the lepton number conservation gained the status of 
firmly established law of Nature. At present, experiment provides one 
the most stringent bounds on violation of this conservation law \cite{4}.
In fact the present experimental data indicates that each of three lepton
generations has his own conserved lepton (flavor) number: in all processes,
investigated until recently, electron number $L_e$, muon number $L_\mu$
and tau number $L_\tau$ are separately conserved. This is in a sharp 
contrast with hadron sector, there only the global baryon number is 
conserved and individual quark flavors are mixed through their couplings
to the electroweak Higs field. In the Standard Model this situation is 
simply postulated by assuming that there are no right-chirality neutrinos
and therefore neutrinos are massless. This is no more true in various
extensions of the Standard Model \cite{5}. But if neutrinos are massive,
the same kind of flavor mixing is expected in lepton sector, as is 
observed for quarks. It is amusing that lepton flavor violation of this
type (neutrino oscillations) were predicted \cite{6} before any notion
of quarks ever emerged in physics, but it took four decades to get at
last strong experimental indications \cite{7,8} that this phenomenon 
actually takes place. Yet, if the Super-Kamiokande result is really due 
to $\nu_\mu \leftrightarrow \nu_\tau$ (or $\nu_\mu \leftrightarrow \nu_s$)
oscillations, there is some mystery why neutrino masses are so miserably
small. The mystery which maybe lead to the mirror world \cite{9}.

Neutrino mixing, if present, can trigger other lepton flavor violating
processes. Search for such kind of decays began half a century ago 
\cite{10}, before lepton flavor was even invented! No positive signal
was ever seen, however. Very small neutrino masses explain this negative 
result: flavor changing transitions generated by neutrino mixing turned
out to be proportional to $\frac{\Delta m_\nu^2}{M_W^2}$ due to lepton
analog of GIM mechanism, and so are extremely suppressed {\cite 4}. However,
neutrino mixing is not the only possibility to have lepton flavor changing
transitions, and many extensions of the Standard Model exist \cite{4}
which allow such transitions with much bigger expected rate.

The tremendous success of the Standard Model initiated a wide spread of 
the philosophy that the only respectable symmetries are gauge symmetries.
Lepton number symmetries are global $U(1)$ symmetries which is deadly
wrong in light of this philosophy: any attempt to make them local will
lead to the immediate conflict with the fact that the corresponding
leptonic photons and mediated by them long range interactions were never
observed. But global symmetries, whatever their origin would be, is
commonly believed to be violated by quantum gravity. The reasoning is
simple \cite{11}. Global charges, like lepton flavor, can be eaten by 
black holes, which subsequently may evaporate. Therefore nonperturbative
quantum gravity effects with formation and evaporation of ``virtual black
holes'' can trigger global symmetry violation. The same kind of effect
can occur if wormholes exist which may take global charges from our 
universe to some other one \cite{12}. Note, however, that the violation
of global symmetries due to wormhole effects could be suppressed by
$e^{-S}$, where $S$ is the wormhole action which may eat the global charge
\cite{11}.

On the contrary, local symmetries, it is believed, can not be violated by 
quantum gravity. The reason is not that black holes can not eat local 
charges like electric one. Certainly they do eat them with the same
appetite as the global ones. But since the local charge is given by 
a surface integral at spatial infinity, because of the Gauss law,
its conservation can not be affected by localized process in space-time,
as is the black-hole evaporation \cite{11,12}. This means that the
black holes fattened by local charge will eventually form charged extreme
black holes which do not evaporate any further \cite{11}.

But lepton flavor symmetries are not local symmetries. So they are 
expected to be violated at list on quantum gravity scale $M_P\approx 
10^{19} \; GeV$. What this means for low energy physics at electroweak
scale? Usually it is assumed that effective lepton flavor violating
unrenormalizable vertexes will be generated in the low energy effective
Lagrangian, which are scaled by inverse powers of $M_P$. In spite of 
a huge difference between electroweak and the Planck scales, such high 
dimensional operators can lead to the observable low energy effects 
\cite{12}.    

Recently a very bold idea was pushed forward \cite{14,15} that actually
quantum gravity scale may be as low as TeV, if some extra spatial 
dimensions are compactified at scales large compared to the electroweak
scale. If $R$ stands for the compactification radius of these large
extra dimensions and $M_Q$ -- for the quantum gravity scale in the 
underlying high dimensional theory, the following relation was obtained 
in \cite{15}
\begin{equation}
M_P^2=R^nM_Q^{n+2}
\label{eq1} \end{equation}
\noindent where $n$ is the number of extra large dimensions. Assuming
$M_Q\sim 1 ~TeV$, this relation implies
\begin{equation}
R\sim 10^{32/n-19} ~m \; .
\label{eq2} \end{equation}
\noindent Already $n=2$ leads to $R\approx 1~mm$ -- just the distance 
beneath which gravitational interactions have not been yet tested 
experimentally. It is further assumed that usual Standard Model fields
do not directly feel these extra dimensions, because they are stuck
on a wall (``3-brane'') in the higher ($\ge 4+n$) dimensional space-time.

Although crazy at first sight, this idea seems not to be in conflict
with the present experimental data, and, on the contrary, provides
interesting phenomenological implications, as was revealed by ever 
increasing number of investigations [15$\div$33]. Moreover, the picture
of ordinary particles confined on the brane, because they correspond to
endpoints of open strings attached to the brane, while gravity, 
represented by closed strings, propagating
in the bulk, was realized in the framework of several string models \cite{34}.

Note however, that it has recently been pointed out \cite{35} that quantum 
gravity effects can lead to an effective energy-dependent light velocity
in vacuum, some kind of ``intrinsic birefringence'' of quantum space-time.
It was further indicated that $\gamma$-ray bursts, which have short duration,
high energies and cosmological origin, can be used to test this prediction.
Recent analysis \cite{36} of TeV $\gamma$-ray flare from active galaxy
Markarian 421 can be interpreted \cite{36} as lower bound on quantum gravity
energy scale $4\cdot 10^{16} GeV$. But it is too premature to conclude that
this experimental result excludes low scale quantum gravity scenario 
advocated in \cite{15}. Clearly a detailed study of quantum gravity 
birefringence in the framework of large extra dimensions model is, however,
desirable.

Nevertheless, if this fascinating $M_Q\sim TeV$ scale quantum gravity 
scheme has something to do with reality, flavor violating effective 
operators are expected at energies much less than $M_Q$. Because experiment
indicates severe restrictions on such operators, this can be quite an ordeal
for the new theory. This question was addressed in \cite{19} and \cite{32}.
The results of \cite{32} indicate that the ``democratic'' maximal mixing
between all three families is excluded and only two of them can have 
significant mixing. 

Usually, the present experimental data is more restrictive for lepton flavor
mixing  between the first and second generations. Data involving the third
generation are scarce. This leaves more room for imagination. In this note
we assume that significant lepton flavor mixing between the third and second
families is generated by low scale quantum gravity (or by some other, 
yet unknown, physics beyond the Standard Model. The Super-Kamiokande data
\cite{7} can be a hint towards such phenomenon) and consider the 
phenomenological implications of the corresponding effective 4-fermion
operators. Our primary interest will be such lepton flavor violating decays
as $Z \to \mu \tau$ \cite{37} and $J/\Psi, \Upsilon \to \mu \tau$ \cite{38}.
These decays escaped much attention, on the contrary to the lepton flavor 
violating decays involving the first and second generation leptons \cite{4}. 

%\section{Dimension-5 operator and $\tau \to \mu \gamma$ decay}

\section{Dimension-6 operators and $Z\to \mu \tau$ decay}
As our aim is to estimate how significant quantum gravity induced 
lepton-flavor violation could be in lucky circumstances, we do not consider
all possible lepton-flavor violating 4-fermion operators. Which of them
(if any) actually survive in the low energy effective action depends on
family-symmetry structure of the underlying theory \cite{32} and can not
be unambiguously guessed from the present experimental data.

Instead, we will focus on some specific 4-fermion operators, which are
not excluded at present, and which can lead, as we will see, to sizeable
lepton-flavor violation effects.

Suppose the following dimension-6 operators are somehow generated
\begin{eqnarray} &
\Delta{\cal L}=\xi_S\Delta{\cal L_S}+\xi_V\Delta{\cal L_V} & 
\nonumber \\ &
\Delta{\cal L_S}=\frac{4\pi\alpha_{S}}{\Lambda^2} \left [
(\bar \tau_R \tau_L)(\bar \tau_L \mu_R)+
(\bar \tau_L \tau_R)(\bar \tau_R \mu_L) \right ] & \\ &
\Delta{\cal L_V}=\frac{4\pi\alpha_{V}}{\Lambda^2} \left [
(\bar \tau_R \gamma_\mu\tau_R)(\bar \tau_R\gamma^\mu \mu_R)+
(\bar \tau_L \gamma_\mu\tau_L)(\bar \tau_L\gamma^\mu \mu_L)\right ] & 
\; , \nonumber
\label{eq3} \end{eqnarray}
\noindent where $\xi_S,\; \xi_V=\pm 1$ and $\Lambda \sim 1 TeV$ is 
the new physics scale (quantum gravity?), which generates lepton-flavor 
violation.

These 4-fermion interactions contribute into  $Z\to \mu \tau$ decay
amplitude through the following one-loop diagrams:

\input FEYNMAN
\begin{picture}(20000,18000)
\drawline\gluon[\E\REG](6000,13000)[4]
\global\advance\pfrontx by -800
\global\advance\pfronty by -200
\put(\pfrontx,\pfronty){$Z$}
\global\advance\pbackx by 2000
\put(\pbackx,\pbacky){\circle{4000}}
\global\advance\pbacky by 2000
\drawarrow[\E\ATBASE](\pbackx,\pbacky)
\global\advance\pbacky by -4000
\drawarrow[\W\ATBASE](\pbackx,\pbacky)
\global\advance\pbacky by 2000
\global\advance\pbackx by 2000
\put(\pbackx,\pbacky){\circle*{500}}
\global\advance\pbackx by 500
\put(\pbackx,\pbacky){\circle*{500}}
\drawline\fermion[\NE\REG](\pbackx,\pbacky)[4000]
\global\advance\pbackx by 200
\put(\pbackx,\pbacky){$\tau$}
\drawarrow[\NE\ATBASE](\pmidx,\pmidy)
\drawline\fermion[\SE\REG](\pfrontx,\pfronty)[4000]
\global\advance\pbackx by 200
\put(\pbackx,\pbacky){$\bar \mu$}
\drawarrow[\NW\ATBASE](\pmidx,\pmidy)
\global\advance\pbackx by 3000
\put(\pbackx,13000){''s-channel'' diagram}
\drawline\gluon[\E\REG](6000,4000)[4]
\global\advance\pfrontx by -800
\global\advance\pfronty by -200
\put(\pfrontx,\pfronty){$Z$}
\global\advance\pbackx by 2000
\put(\pbackx,\pbacky){\circle{4000}}
\global\advance\pbacky by 2000
\drawarrow[\E\ATBASE](\pbackx,\pbacky)
\global\advance\pbacky by -4000
\drawarrow[\W\ATBASE](\pbackx,\pbacky)
\global\advance\pbacky by 2500
\global\advance\pbackx by 2000
\put(\pbackx,\pbacky){\circle*{500}}
\drawline\fermion[\NE\REG](\pbackx,\pbacky)[4000]
\global\advance\pbackx by 200
\put(\pbackx,\pbacky){$\tau$}
\global\advance\pfronty by -500
\put(\pfrontx,\pfronty){\circle*{500}}
\drawline\fermion[\SE\REG](\pfrontx,\pfronty)[4000]
\global\advance\pbackx by 200
\put(\pbackx,\pbacky){$\bar \mu$}
\drawarrow[\NW\ATBASE](\pmidx,\pmidy)
\global\advance\pbackx by 3000
\put(\pbackx,4000){''t-channel'' diagram}
\end{picture}

Let us estimate how big could be the branching ratio
$Br(Z\to\mu\tau)$. The corresponding neutral current matrix element is
conserved $q_\mu J^\mu=0$ ($q_\mu$ being the $Z$-boson 4-momentum) in the
approximation when final lepton masses are neglected, so it can be 
parameterized as follows
\begin{eqnarray} &
J_\mu= & \\ &
=\bar u(p_\tau)\left [ (F_{E0}+\gamma_5 F_{M0})\gamma^\nu (g_{\mu\nu}-
\frac{q_\mu q_\nu}{q^2})+(F_{M1}+\gamma_5 F_{E1})i\sigma_{\mu\nu}\frac{q^\nu}
{m_\tau} \right ] v(p_\mu) . \nonumber
\label{eq4} \end{eqnarray}
\noindent Then the standard consideration leads to the decay width
\begin{eqnarray}
\Gamma(Z\to\bar\mu\tau)=\frac{M_Z}{24\pi}\left [ 2(|F_{E0}|^2+
|F_{M0}|^2)+\frac{M_Z^2}{m_\tau^2}(|F_{E1}|^2+|F_{M1}|^2)\right ] & \; .
\label{eq5} \end{eqnarray}
The above given one-loop diagrams contribute only into the $F_{M0}$ 
form-factor (for the specific form of 4-fermion interactions we are
considering). To deal with ultraviolet divergences, inherent for these 
diagrams, we use, as suggested in \cite{39,40}, dimensional regularization
and the $\overline{MS}$ renormalization scheme, choosing the renormalization
scale $\mu=\Lambda$. We find immediately that the scalar-scalar interaction
${\cal L}_S$ contributes only through ``t-channel'' diagram, while, on
contrary, vector-vector interaction ${\cal L}_V$ contributes only through
``s-channel'' diagram. Normalizing over $\Gamma(Z\to\mu^+\mu^-)$ and
summing over two charged modes $Z\to\mu^+\tau^-$ and $Z\to\mu^-\tau^+$,
we get
\begin{eqnarray}
\frac{\Gamma(Z\to\mu^\pm\tau^\mp)}{\Gamma(Z\to\mu^+\mu^-)}=
\frac{(2\xi_V\alpha_V-\xi_S\alpha_S)^2}{9\cdot (4\pi)^2}~
\left (\frac{M_Z}{\Lambda}\right )^4 \left [ \left (
\ln{\frac{M_Z^2}{\Lambda^2}}-\frac{5}{3} \right )^2 +\pi^2 \right ].
\label{eq6} \end{eqnarray}
\noindent For $\xi_V=1, \; \xi_S=-1, \; \Lambda \sim 1 TeV$ and strong
coupling regime $\alpha_S, \alpha_V \sim 1$, the last equation implies
$$\frac{\Gamma(Z\to\mu^\pm\tau^\mp)}{\Gamma(Z\to\mu^+\mu^-)}\approx
5\cdot 10^{-5}.$$
\noindent In terms of branching ratio this translates into
$Br(Z\to\mu^\pm\tau^\mp)\approx 1.6\cdot 10^{-6}$ -- about order
of magnitude below of present upper limit, but not too small to move it
beyond the reach of future experiments.  

The same type of loop-diagrams induce lepton-flavor changing electromagnetic
current $\gamma \to\mu\tau$. 
%(the identity of a particle is changed through
%the emission and absorption of a photon! Times had changed since 1948. Today
%nobody is shocked by such statement). 
Now equation (4) gives the general 
form for the matrix element of such current \cite{41,4}. Note, however, that
the one-loop diagrams, we are interested in, gives form factors which vanish
for a real photon $q^2=0$, so there are no contributions into $\tau\to\mu
\gamma$ decay.

One of the consequences of the $\gamma \to\mu\tau$ transition is the 
induced $V\to\mu\tau$ decay, where $V=J/\Psi$ or $\Upsilon$. In general
we have
$$\frac{\Gamma(V\to\mu^\pm\tau^\mp)}{\Gamma(V\to\mu^+\mu^-)}=
(1-r)^2\zeta \; , $$
\noindent where $r=m_\tau^2/M_V^2$ and
$$\zeta=(2+r)\left(|F_{E0}(M_V^2)|^2+|F_{M0}(M_V^2)|^2\right) + \left (
2+\frac{1}{r}\right )\left(|F_{E1}(M_V^2)|^2 \right .
+$$ $$+\left . |F_{M1}(M_V^2)|^2\right)
+6~Re \left [F_{E1}(M_V^2)F_{M0}^*(M_V^2)-F_{E0}(M_V^2)F_{M1}^*(M_V^2)
\right ] \; . $$

Unfortunately the 4-fermion interactions (3) turn out to give negligible
contributions to these decays. For $\Upsilon\to\mu\tau$ we get
$$\frac{\Gamma(\Upsilon\to\mu^\pm\tau^\mp)}{\Gamma(\Upsilon\to\mu^+\mu^-)}=
\frac{\alpha}{18\pi}(4\xi_V\alpha_V-\xi_S\alpha_S)^2\left ( \frac{
M_\Upsilon}{\Lambda}\right )^4 \left[  \left (
\ln{\frac{M_\Upsilon^2}{\Lambda^2}}-\frac{5}{3} \right )^2 +\pi^2 \right ].
$$
\noindent That is approximately $3\cdot 10^{-9}$.
While for the $J/\Psi\to\mu\tau$ decay ( $J/\Psi\to\tau\tau$ is
kinematically forbidden for on-shell $\tau$-s, so the form factors do not
have an imaginary part)
$$\frac{\Gamma(J/\Psi\to\mu^\pm\tau^\mp)}{\Gamma(J/\Psi\to\mu^+\mu^-)}=$$
$$=\frac{\alpha}{36\pi}(1-r)^2(2+r)(4\xi_V\alpha_V-\xi_S\alpha_S)^2
\left ( \frac{
M_{J/\Psi}}{\Lambda}\right )^4   \left (
\ln{\frac{M_{J/\Psi}^2}{\Lambda^2}}+F(r) \right )^2 . $$
\noindent Now r is not   small and we should use the full
loop-integral function
$$F(r)=\ln{r}-4r-\frac{5}{3}+2(1+2r)\sqrt{4r-1}~arctg{\left (
\frac{1}{\sqrt{4r-1}} \right )}\; , $$
\noindent but this doesn't help much, anyway -- the above given ratio
is not likely to exceed $3\cdot 10^{-11}$.

\section{More Dimension-6 operators and $V\to \mu \tau$ decays}
The above considered lepton-flavor changing decays have small rates partly
because they are caused by one-loop effects. We can avoid loop suppression
factor $(4\pi)^{-2}$ in the decay amplitude if manage to find such 4-fermion
operators which allow lepton-flavor violating decays at tree level. Is this
possible? The answer is yes. If we do not bound ourselves by purely leptonic
4-fermion operators, the following dimension-6 operators can do the job for
$V\to\mu\tau$ decays:
\begin{eqnarray} &
\Delta{\cal L}=\xi_c\Delta{\cal L}_c+\xi_b\Delta{\cal L}_b & 
\nonumber \\ &
\Delta{\cal L}_c=\frac{4\pi\alpha_{c}}{\Lambda^2} \left [
(\bar c_R \gamma_\mu c_R)(\bar \tau_R\gamma^\mu \mu_R)+
(\bar c_L \gamma_\mu c_L)(\bar \tau_L\gamma^\mu \mu_L)\right ] & \\ &
\Delta{\cal L}_b=\frac{4\pi\alpha_{b}}{\Lambda^2} \left [
(\bar b_R \gamma_\mu b_R)(\bar \tau_R\gamma^\mu \mu_R)+
(\bar b_L \gamma_\mu b_L)(\bar \tau_L\gamma^\mu \mu_L)\right ] & \; .
\nonumber \label{eq7} \end{eqnarray}
\noindent One more benefit from such operators is that the $\alpha$
(electromagnetic) suppression is also overcome in the $V\to\mu\tau$
amplitude. 

Of course we should take care about the fact that the quarks are not free 
but bound in the $V$ vector meson. Nonrelativistic approximation looks, 
however,
reasonable for heavy enough $c$ and $b$ quarks. So simplification can be
achieved by using the fact that in the weak binding limit the state 
vector of the $V(Q \bar Q)$ vector meson can
be represented as a superposition of the free quark-antiquark states
\cite{42}:
$$ 
|V(\vec P,\epsilon)>= $$ $$
=\sqrt{2M_V}\int\frac{d\vec p}{(2\pi)^3}
\sum\limits_{ms\bar s}C^{1m}_{s\bar s}~\epsilon \cdot \epsilon^*_{(m)}
\varphi_{V}(\vec P)|Q[\frac{m_Q}{M_V}\vec P+\vec p,s]
\bar Q[\frac{m_Q}{M_V}\vec P-\vec p,\bar s]>  . $$
\noindent
We use $<\vec P\,^\prime|\vec P>=(2\pi)^3 2E\delta (\vec P\,'-\vec P)$ 
normalization for the meson state vectors while
$<\vec p\,^\prime|\vec p>=(2\pi)^3~\frac{E}{m}~\delta(\vec p\,'-\vec p)$
one for the state vectors of quark (or antiquark) with mass m. 
$\epsilon_{(-)}, \; \epsilon_{(0)}$ and $\epsilon_{(+)}$ are three 
independent polarization 4-vectors for the vector mesons. 
$C^{jj_z}_{s\bar s}$ are the usual 
Clebsh-Gordan coefficients that couple $s$ and $\bar s$ quark and
antiquark spins into the vector-meson spin and polarization $\epsilon_\mu$.
 At last,
$|Q[\vec p_1,s]\bar Q[\vec p_2,s]>=a^+_s (\vec p_1) b^+_{\bar s}
(\vec p_2)|0>$, $a^+$ and $b^+$ being the quark and antiquark creation
operators. Note that our normalization convention indicates the
following anticommutation relations
$$
\{a_s(\vec p),a^+_{s'}(\vec p\,') \}=\{b_s(\vec p),b^+_{s'}
(\vec p\,') \}=(2\pi)^3\frac{E}{m}\delta (\vec p\,'-\vec p) \; .
$$
\noindent Now all what is needed is to decompose quark field operators
in the interaction Lagrangian $\Delta{\cal L}$ in terms of creation
and annihilation operators and use the above given anticommutation relations
while calculating lepton-flavor changing neutral current matrix element,
along with the 
nonrelativistic approximation $E=\sqrt{m^2 +\vec p ^2}\approx m$. As a 
result, we get (in the vector-meson rest frame; $\Psi(0)$ is the coordinate 
space wave function at the origin and $N_c=3$ is for color), 
for example
$$<0|(\bar c_L \gamma_\mu c_L)|J/\Psi>=$$
$$=\sqrt{2M_V}N_c\Psi(0)
\sum\limits_{ms\bar s}C^{1m}_{s\bar s}~\epsilon \cdot \epsilon^*_{(m)}
\bar v_{(\bar s)}(0)\gamma_\mu \frac{1}{2}(1-\gamma_5)u_{(s)}(0). $$
\noindent Now it is straightforward to get the following decay width
\begin{equation}
\Gamma(J/\Psi \to \mu^\pm \tau^\mp)=8\pi\alpha_c^2
\left (\frac{M_{J/\Psi}}{\Lambda}\right )^4~\frac{|\Psi(0)|^2}{M_{J/\Psi}^2}
, \label{eq8} \end{equation}
\nonumber while for the $\Gamma(J/\Psi \to \mu^+ \mu^-)$ we recover the
well known result
$$
\Gamma(J/\Psi \to \mu^+ \mu^-)=\frac{16\pi\alpha^2e_c^2}
{M_{J/\Psi}^2}|\Psi(0)|^2 \; , $$
\noindent where $e_c=2/3$ stands for the $c$-quark electric charge (in terms
of positron charge).

Analogous consideration holds for the $\Upsilon$-meson decay. Both cases
can be unified by the formula
\begin{equation}
\frac{\Gamma(V \to \mu^\pm \tau^\mp)}{\Gamma(V \to \mu^+ \mu^-)}=
\frac{1}{2e_Q^2}\left (\frac{\alpha_Q}{\alpha} \right )^2
\left (\frac{M_V}{\Lambda} \right )^4.
\label{eq9} \end{equation} 
Assuming once more strong coupling $\alpha_Q \sim 1$ and $\Lambda\sim 1TeV$,
this formula predicts big enough branching ratios:
$$ Br(J/\Psi \to \mu^\pm \tau^\mp)\approx 1.2\cdot 10^{-7}, \; \;
Br(\Upsilon \to \mu^\pm \tau^\mp)\approx 2\cdot 10^{-5}. $$
\noindent Having in mind near future high luminosity meson factories,
such rates are clearly observable, aren't they?

Note that (7) interactions will contribute also into $Z\to\mu\tau$ decay.
So the branching ratio for this decay may be even somewhat higher, than
given above, if all contributions interfere constructively.

\section{Peculiarity of $J/\Psi \to \mu \tau$ decay}
It is a pity that the small mass of $J/\Psi$ meson makes it less sensitive
to lepton-flavor violation originated from a new physics at higher scales,
because this very smallness of its mass prevents it from decaying into
$\tau^+\tau^-$, so considerably reducing background. Anyhow this peculiarity
of $J/\Psi \to \mu \tau$ decay may be of practical interest. In this section
we take a closer look to a possible signature of this decay mode.

Let us examine secondary muon spectrum from the final state $\tau$ decay
(because of short lifetime, this final state  $\tau$ is not directly
detectable). In the $\tau$-lepton rest frame $\mu$-spectrum is
\begin{equation}
\rho^*(x^*)=\frac{\Delta N}{\Delta x^*}=16x^{*2}(3-4x^*) \; ,
\label{eq10} \end{equation}
\noindent where $x^*=\frac{E_\mu^*}{m_\tau}$, $E^*_\mu$ being the muon
energy in this frame.

\begin{figure}[htb]
  \begin{center}
\mbox{\epsfig{figure=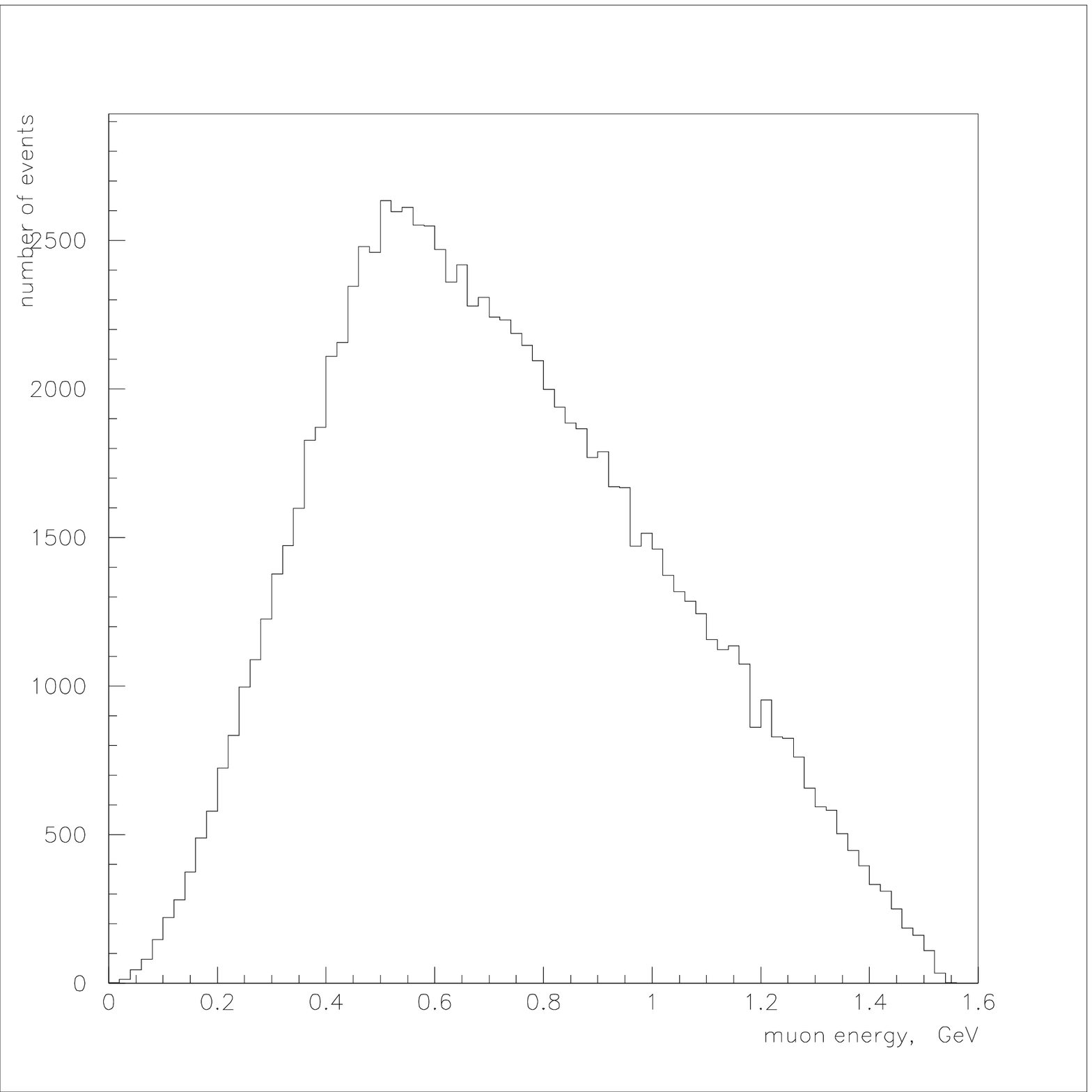
                               ,  height=15.0cm}}
   \end{center}
\caption {Secondary muons energy spectrum in $J/\Psi$ rest frame}
\label{Fig1}
\end{figure}

In the laboratory frame ($J/\Psi$-meson rest frame) the muon energy
$E_\mu$ is given by Lorentz transformation
\begin{equation}
E_\mu=\gamma_\tau E^*_\mu (1+\beta_\tau \cos{\theta^*} ) \; ,
\label{eq11} \end{equation}
\noindent where 
$$ \cos{\theta^*}=\frac{\vec{p^*_\mu}\cdot \vec{p_\tau}}{|\vec{p^*_\mu}|
|\vec{p_\tau}|} $$
\noindent and we neglect the muon mass. But the $\tau$-lepton energy in 
the $J/\Psi \to \mu \tau$ decay is $E_\tau=\frac{1+r}{2}M_{J/\Psi}$,
where $r=\frac{m_\tau^2}{M_{J/\Psi}^2}$. So
$$\gamma_\tau=\frac{E_\tau}{m_\tau}=\frac{1+r}{2}~\frac{M_{J/\Psi}}{m_\tau}
\; \; \; {\rm and} \; \; \; \beta_\tau=\sqrt{1-\frac{1}{\gamma_\tau^2}}=
\frac{1-r}{1+r} \; . $$
\noindent Therefore (11) can be rewritten as 
$$E_\mu=\frac{1+r}{2}~\frac{M_{J/\Psi}}{m_\tau}\left (1+\frac{1-r}{1+r}
\cos{\theta^*} \right )E^*_\mu$$
\noindent or, introducing the muon energy fraction $x=\frac{E_\mu}
{M_{J/\Psi}}$ in the laboratory frame,
\begin{equation}
x=\frac{x^*}{2} \left [ 1+r+(1-r)\cos{\theta^*} \right ] \; .
\label{eq12} \end{equation}
\noindent For a given $x$, $~x^*$ may change from $x$ (when $\cos{\theta^*}=
1$) to ${\rm min}\left (\frac{1}{2},\frac{x}{r}\right )$ 
(when $\cos{\theta^*}=-1$
or $x^*$ reaches its maximal value of $\frac{1}{2}$). Besides, because
$$\cos{\theta^*}=\frac{1}{1-r}\left[ \frac{2x}{x^*}-1-r \right ] \; , $$
\noindent we have
$$\Delta \cos{\theta^*}=\frac{2\Delta x}{(1-r)x^*} \; . $$
\noindent But the distribution of muons over $\Delta \cos{\theta^*}$ is
isotropic (if the initial $J/\Psi$ is not polarized).
So from $\rho^*(x^*)\Delta x^*$ secondary muons only
$$\Delta N=\rho^*(x^*)\Delta x^*\frac{\Delta \cos{\theta^*}}{2}=
\frac{\rho^*(x^*)\Delta x^*}{(1-r)x^*}\Delta x$$
\noindent will have the energy fraction in the limits $x\pm \Delta x$.
Therefore we get the following distribution
$$\rho(x)\equiv \frac{\Delta N}{\Delta x}=
\frac{\Theta(r-2x)}{1-r}\int\limits_x^{\frac{x}{r}} 
\frac{\rho^*(x^*)}{x^*}~dx^* +
\frac{\Theta(2x-r)}{1-r}\int\limits_x^{\frac{1}{2}} 
\frac{\rho^*(x^*)}{x^*}~dx^* $$
\noindent or, substituting (10) and integrating,
$$\rho(x)=
\left [ \frac{8(1+r+r^2)(9r-8x)}{3r^3}-24 \right ]x^2\Theta(r-2x)+
$$
\begin{equation}
+\frac{2(1-2x)(5+10x-16x^2)}{3(1-r)}\Theta(2x-r). 
\label{eq13} \end{equation}
This spectrum is illustrated on Fig.\ref{Fig1}. Note that energy of primary
muon (not from the $\tau$, but from the initial $J/\Psi$ decay) is
$(M_{J/\Psi}^2-m_\tau^2)/(2M_{J/\Psi})\approx 1.04 GeV$, while muons from
the background  $J/\Psi \to \mu^+ \mu^-$ decay have equal energies of 
$1.55GeV$ each. As we see muons from the $J/\Psi \to \mu \tau, \;
\tau \to \mu\nu\bar\nu$ cascade are considerably softer than muons
from the most important background. Therefore we expect rather clean 
experimental signature for the $J/\Psi \to \mu \tau$ lepton-flavor violating
decay.
  
\section{Conclusions}
As we have seen, the lepton flavor violating decays, considered above, can
proceed at an observable level, if the underlying 4-fermion operators
are generated via recently proposed low scale quantum gravity effects,
unless such operators are forbidden (or suppressed) by some symmetry principle. 
Of course, the same 4-fermion operators could be generated also by some
other new physics at a TeV scale, not necessarily quantum gravity. So these
decays are not unambiguous signals of the large extra dimensions. Nevertheless,
the experimental study of the $Z \to \mu \tau$ and $J/\Psi, \Upsilon 
\to \mu \tau$ decays is of considerable interest, because the existing or
near future facilities will allow to place a stringent limit on the
lepton flavor violation in these channels, or discover this interesting 
phenomenon.

\section*{Acknowledgments}
The author is grateful to Z.~Berezhiani for valuable comments.
He also acknowledges useful conservations with V.~I.~Telnov about 
lepton-flavor
violation, which ultimately lead to this investigation.

\newpage
%                  Figures

\end{document}